\documentclass[prl,twocolumn]{revtex4}
\usepackage{graphicx}
\usepackage{float}
\usepackage{amsfonts}
\usepackage{makeidx}
\usepackage{graphicx}
\usepackage{amsmath,amssymb,graphics,epsfig,color,times,bbm}

\begin{document}

\title{Continuous variable entanglement enhancement and manipulation by a sub-threshold type-II optical parametric amplifier}

\author{Yana Shang, Xiaojun Jia$^{*}$, Yumei Shen, Changde Xie, Kunchi Peng}
\address{State Key Laboratory of Quantum Optics and Quantum Optics Devices,\\
Institute of Opto-Electronics, Shanxi University, Taiyuan, 030006,
P. R. China}

\address{$^*$Corresponding author: jiaxj@sxu.edu.cn}

\begin{abstract}We experimentally demonstrate that the quantum entanglement
between amplitude and phase quadratures of optical modes produced
from a non-degenerate optical parametric amplifier (NOPA) can be
enhanced and manipulated phase-sensitively by means of another NOPA.
When both NOPAs operate at de-amplification, the entanglement degree
is increased at the cavity resonance of the second NOPA. When the
first NOPA operates at de-amplification and the second one at
amplification, the spectral features of the correlation variances
are significantly changed. The experimental results are in good
agreement with the theoretical expectation.
\end{abstract}

\maketitle 

The entangled states of light with amplitude and phase quadrature
quantum correlations have been extensively applied in continuous
variable (CV) quantum information\cite{Bra,Reid,Li,Aur}. It is
significant to enhance and manipulate quantum entanglement of
optical entangled states for realizing high quality information
processing and achieving long distance quantum communication. The
phase-sensitive responses of quantum states of light through a
degenerate optical parametric amplifier (DOPA) has been
theoretically studied by Agarwal\cite{Aga}. The phase-sensitive
manipulation of a squeezed vacuum field in a DOPA\cite{Zhang} and
the low-noise phase-insensitive amplification of a CV quantum
state\cite {And,Poo} have been experimentally demonstrated. Besides
DOPAs, nonclassical optical fields, such as squeezed states and
entangled states, can be directly generated from a continuous
type-II non-degenerate optical parametric amplifier (NOPA) operated
below and above the oscillation threshold,
respectively\cite{Li,Aur,Vil,Jing,Kel}. Recently, it has been
theoretically proved that the entanglement features of
Einstein-Podolsky-Rosen (EPR) optical beams can be phase-sensitively
manipulated by another NOPA\cite{Chen}. Here, we present the first
experimental demonstration on the phase-sensitive manipulation of
EPR entangled states. Two NOPAs are used in the experiment, one for
the generation of EPR beams and another one for entanglement
manipulating. The quantum entanglement between amplitude and phase
quadratures of EPR beams produced from the first NOPA are enhanced
when both NOPAs operate at de-amplification. When the first NOPA
operates at de-amplification and the second one at amplification,
the spectral features of the correlation variances are significantly
changed. The experimental results demonstrate that the quantum
correlation features of an optical entangled state can be
manipulated by controlling the relative phase between the pump field
and the injected fields of the second NOPA, which are in good
agreement with the theoretical expectation in Ref.\cite{Chen}.

\begin{figure}
\centerline{
\includegraphics[width=3in]{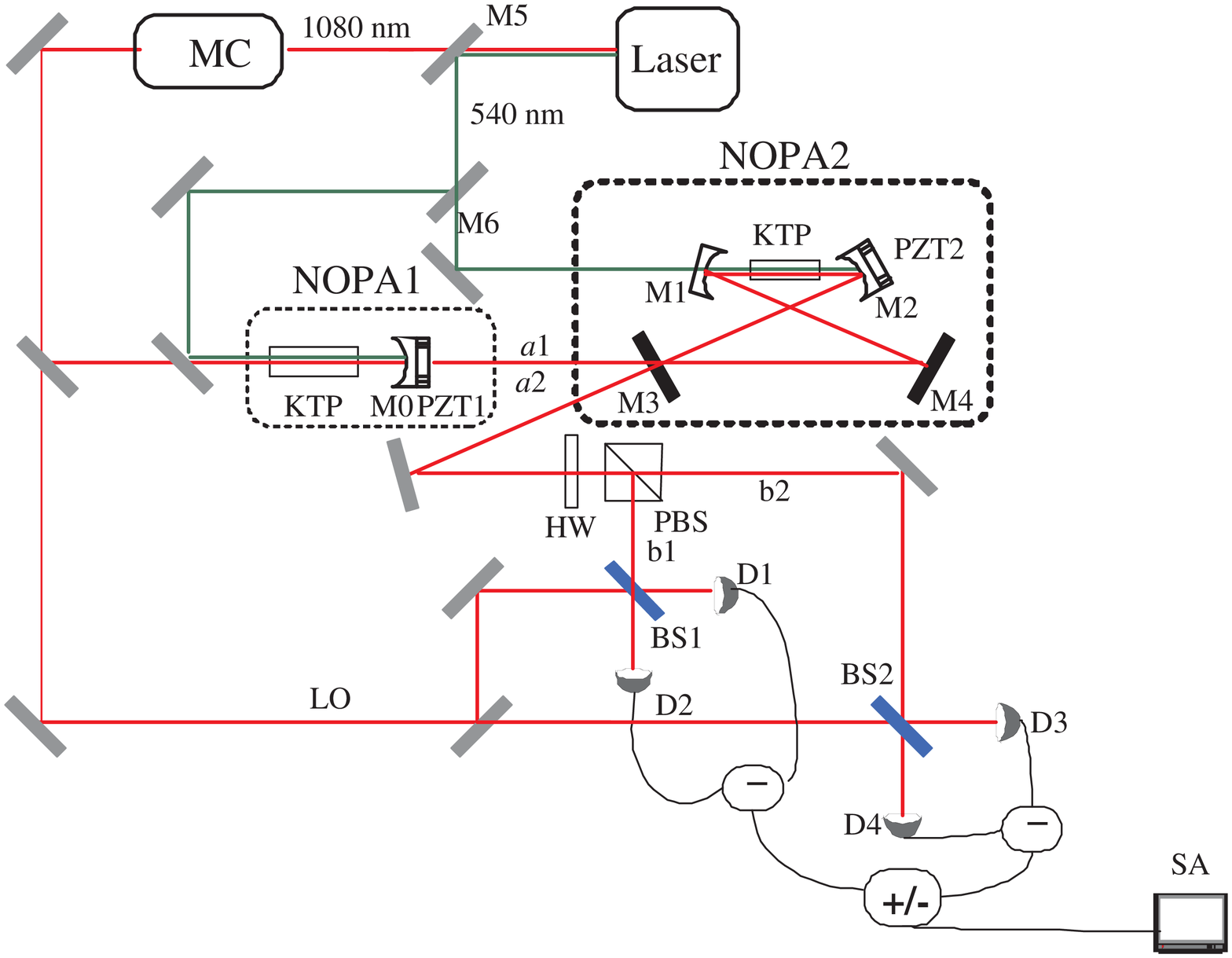}
} \vspace{0.1in}
\caption{(Color online) Experimental setup. laser, Nd:YAP/KTP laser
source; HW, $\lambda /2$ wave plate; PBS, polarizing beam splitter;
BS1-2, 50/50 beam splitter; D1-4, ETX500 InGaAs photodiode
detectors; $+/-$, positive/negative power combiner; MC, mode
cleaner; M0-6, different mirrors (see text for detail); PZT, piezo
electric transducer; SA, spectrum analyzer. \label{Fig1} }
\end{figure}

The experimental setup is shown in Fig.1. A home-made continuous
wave frequency-doubled and frequency-stabilized Nd:YAP/KTP
(Nd-dropped YAlO$_3$ /KTiOPO$_4$) laser with both harmonic (540 nm)
and subharmonic (1080 nm) outputs serves as the light source for the
pump, signal and local oscillator (LO)\cite{Li}. The first NOPA
(NOPA1) is in a semi-monolithic F-P configuration consisting of an $\alpha $%
-cut type-II KTP crystal with 10 mm length and a concave mirror (M0)
with 50 mm radius of curvature. The KTP crystal implements the
non-critical phase matching frequency-down-conversion of the pump
field\cite{Ou}. The front face of the crystal is coated to be used
as the input coupler (the transmission 95.2\% at 540 nm and 0.2\% at
1080 nm), M0 serves as the output coupler for the EPR beam at 1080
nm (the transmission of 3.2\% at 1080 nm), which is mounted on a
piezo electric transducer (PZT) to scan or lock actively the cavity
length as needed. The second NOPA (NOPA2) is in a ring configuration
consisting of two concave mirrors with
100 mm radius of curvature (M1 and M2) and two flat mirrors (M3 and M4). An $%
\alpha $-cut type-II KTP\ crystal of 10 mm length with the 1080 nm
and 540 nm dual-band antireflection coated at two end faces is
placed at the middle between M1 and M2. M3 serves as the input and
the output coupler of light at 1080 nm with the transmission of
3.5\% at 1080 nm. The length and the finesse of the cavity for NOPA1
(NOPA2) are 51 mm (557 mm) and 165 (153), respectively. The output
signal and idler beams ($b_1$ and $b_2$) from NOPA2 are separated by
a polarizing-beam-splitter(PBS) and then they are respectively sent
to two balanced homodyne detectors for simultaneously measuring the
noise power spectra of their quadrature components. The measured
noise power are combined by a positive or negative power combiner
(+/-) and then the combined correlation variances of amplitude or
phase quadratures between $b_1$ and $b_2$ are detected by a spectrum
analyzer (SA).

At first, we achieved the double resonance of injected subharmonic
coherent signal and idler in NOPA1 using FM sideband
technique\cite{Li,Ou}. Then the relative phase between the pump
field and the injected signal field is locked to $\pi $, that is to
enforce the NOPA1 operating at de-amplification\cite{Li}. When a
block was inserted between M3 and M4 and the pump light of NOPA2 was
turned off, the output light by NOPA1 was not coupled into NOPA2 and
almost totally was reflected by M3 to PBS. In this case the
quadrature correlation variances of the EPR beams produced by NOPA1
were measured. The measured correlation variances spectra of the
amplitude sum, $\langle \delta
^2(\hat{X}_{a_1}+\hat{X}_{a_2})\rangle $, and the phase difference,
$\langle \delta ^2(\hat{Y}_{a_1}-\hat{Y}_{a_2})\rangle $, both were
$2.4\pm 0.1$ dB below the corresponding shot noise limit(SNL) at the
analysis frequency of $\Omega =3.0MHz$, where $\hat{X}_{a_1}$ and $\hat{X}%
_{a_2}$ ($\hat{Y}_{a_1}$ and $\hat{Y}_{a_2}$) are the amplitude
(phase) quadratures of output modes $a_1$ and $a_2$ by NOPA1,
respectively. It means that the EPR entangled optical field with the
amplitude anti-correlation and the phase correlation were obtained.
During the experiment the pump power and intensity of the injected
signal for NOPA1 are kept at 120 mW(below the oscillation threshold
of 200 mW) and 10mw before the input coupler, respectively. The
power of the output EPR entangled beams is about $52$ $\mu W$.

\begin{figure}
\centerline{
\includegraphics[width=3in]{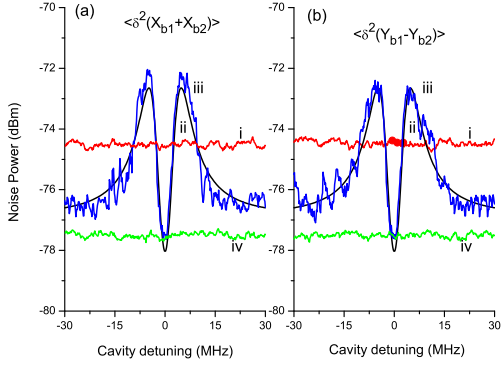}
} \vspace{0.1in}
\caption{(Color online) The noise powers of the correlation
variances of the output beams by NOPA2 operated at de-amplification
versus cavity detuning: (a) for amplitude sum $\langle \delta
^2(\hat{X}_{b_1}+\hat{X}_{b_2})\rangle $; (b) for phase difference
$\langle \delta ^2(\hat{Y}_{b_1}-\hat{Y}_{b_2})\rangle $. Trace i:
SNL; trace ii: theoretically calculated results; trace iii: the
measured variance spectrum with the cavity detuning; trace iv: the
measured variance power with the cavity locked on resonance.
\label{Fig2} }
\end{figure}

Removing the block between M3 and M4 as well as turning on the pump
light of NOPA2, the EPR beams by NOPA1 were injected into NOPA2.
When NOPA2 was operated at de-amplification also, the output signal
and idler modes by NOPA2 were still entangled with the
anti-correlation of amplitude quadratures ($\langle \delta
^2(\hat{X}_{b_1}+\hat{X}_{b_2})\rangle $ $<SNL$) and the
correlation of phase quadratures ($\langle \delta ^2(\hat{Y}_{b_1}-\hat{Y}%
_{b_2})\rangle <SNL$) like the entanglement features of the injected
signals, where $\hat{X}_{b_1}$ and $\hat{X}_{b_2}$ ($\hat{Y}_{b_1}$ and $%
\hat{Y}_{b_2}$) are the amplitude (phase) quadratures of output
modes $b_1$
and $b_2$ by the NOPA2, respectively. The correlation variance spectra of $%
\langle \delta ^2(\hat{X}_{b_1}+\hat{X}_{b_2})\rangle $ and $\langle
\delta ^2(\hat{Y}_{b_1}-\hat{Y}_{b_2})\rangle $ versus the cavity
detuning measured by scanning the length of optical cavity of NOPA2
are shown in Fig.2(a) and (b), respectively. Under the resonance
with zero detuning ($\Delta =0$), both variance of the amplitude sum
(Fig.2(a) trace iii ) and the phase difference (Fig.2(b) trace iii)
are about 3.0 dB below the SNL(trace i). If locking the cavity
length to the resonance point a stable correlation variance of
$3.0\pm 0.1dB$ below the SNL is obtained(trace iv). In this case the
entanglement degree of the output fields by NOPA2 are enhanced about
0.6 dB with respect to that of the injected entangled states. Trace
ii in Fig.2(a) and (b) are the correlation variance spectra of
amplitude sum (a) and phase difference(b) calculated by Eq.(25) in
Ref.\cite{Chen} with the actual parameters of the experimented
system, respectively. Deviating from the resonance with a small
detuning the correlation noises increase rapidly to two maximums of
2.0 dB above the SNL at $\Delta =\pm 4.9MHz$ and then
decrease to the initial correlation degree of the injected EPR beams ($\sim $%
2.4 dB below the SNL) at far detuning.

\begin{figure}
\centerline{
\includegraphics[width=3in]{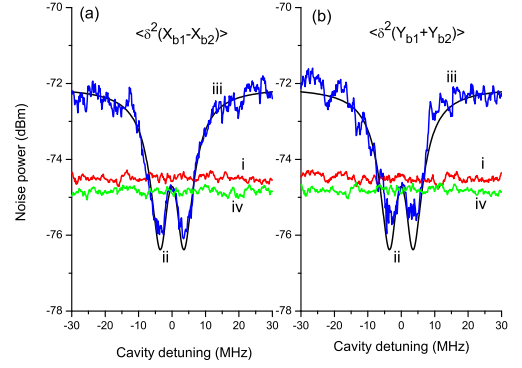}
} \vspace{0.1in}
\caption{(Color online) The noise powers of the correlation
variances of the output beams by NOPA2 operated at amplification
versus cavity detuning: (a) for amplitude difference $\langle \delta
^2(\hat{X}_{b_1}-\hat{X}_{b_2})\rangle $; (b) for phase sum $\langle
\delta ^2(\hat{Y}_{b_1}+\hat{Y}_{b_2})\rangle $. Trace i: SNL; trace
ii: theoretically calculated results; trace iii: the measured
variance spectrum with the cavity detuning; trace iv: the measured
variance power with the cavity locked on resonance. \label{Fig3} }
\end{figure}

However, if NOPA2 is operated at amplification by locking relative
phase between the pump field and the injected EPR beam in phase and
NOPA1 still at de-amplification, the correlation features of
quadratures between the output modes $b_1$ and $b_2$ will be
significantly changed. In this case the noise
powers of $\langle \delta ^2(\hat{X}_{b_1}+\hat{X}_{b_2})\rangle $ and $%
\langle \delta ^2(\hat{Y}_{b_1}-\hat{Y}_{b_2})\rangle $ are not
squeezed. In
contrast with the correlation features of the injected EPR beams with $%
\langle \delta ^2(\hat{X}_{a_1}+\hat{X}_{a_2})\rangle <SNL$ and
$\langle \delta ^2(\hat{Y}_{a_1}-\hat{Y}_{a_2})\rangle <SNL$, the
correlation
variances of the output fields by NOPA2, $\langle \delta ^2(\hat{X}_{b_1}-%
\hat{X}_{b_2})\rangle $ and $\langle \delta ^2(\hat{Y}_{b_1}+\hat{Y}%
_{b_2})\rangle $, become the quantum correlated with noise powers
below the SNL at the cavity resonance and near resonance. Fig.3(a)
and (b) are the
noise power spectra of $\langle \delta ^2(\hat{X}_{b_1}-\hat{X}%
_{b_2})\rangle $ and $\langle \delta
^2(\hat{Y}_{b_1}+\hat{Y}_{b_2})\rangle $ versus the cavity detuning
of NOPA2 at $\Omega =3.0MHz$, respectively. Trace i is the SNL;
trace ii and iii are the noise power spectra calculated by Eq.(25)
in Ref.\cite{Chen} and experimentally measured, respectively; trace
iv is the noise spectra measured when the cavity of NOPA2 is locked
at the resonance . Both correlation variances of amplitude
difference (a) and phase sum (b) at the cavity resonance are 0.4 dB
below the SNL and the minimal variances of 1.4 dB below the SNL
appear at small detuning of $\Delta =\pm 3.5MHz$. Then, the
variances increase to much higher than the SNL at far
detuning where the parametric interaction in NOPA2 no longer exist and thus $%
(\hat{X}_{b_1}-\hat{X}_{b_2})$ and $(\hat{Y}_{b_1}+\hat{Y}_{b_2})$
return to anti-squeezing components of the initially injected modes
$a_1$ and $a_2$.

When NOPA1 and NOPA2 are operated at the same regime (Fig.2) the
identical parametric interaction in NOPA2 will enhance the
entanglement of the injected signal field. However, if NOPA1 and
NOPA2 are operated at the opposite regime (Fig.3), the quantum
correlations produced by the nonlinear process in NOPA2 have the
opposite features with the correlations of the injected signal field
at the resonance and the near resonance. Thus the amplitude
anti-correlation and the phase correlation of the injected field are
changed to the amplitude correlation and the phase anti-correlation
of the output field due to the parametric amplification process in
NOPA2. The shoulders appearing in the spectral shapes just outside
the resonant point are caused by the interference between the pump
field and the subharmonic seed field in NOPA2 in cooperation with
the absorptive and dispersive responses of an optical
cavity\cite{Chen}. Comparing trace ii and iii in Figs.(2) and (3),
we can see that theoretically calculated and experimentally measured
correlation variance spectra are in good agreement except at the
dips of the minimal variances. At these dips the calculated
variances are smaller than measured values, that is perhaps because
some extra instability appears at the suddenly changing points of
the correlation variances which have not be involved in the
theoretical equations.

For conclusion, we experimentally realized the entanglement
enhancement and phase-sensitive manipulation of CV optical entangled
states based on using NOPAs. The experiment provides a simple scheme
to increase and manipulate CV quantum correlations of optical modes
without the need for the difficult technique of single photon
detection\cite{Our}.

Acknowledgement: This research was supported by Natural Science
Foundation of China (Grants Nos. 60736040 and 60608012), NSFC
Project for Excellent Research Team (Grant No. 60821004), National
Basic Research Program of China (Grant No. 2010CB923103).

\end{document}